\documentclass[a4paper]{article}

\usepackage{INTERSPEECH2022}
\usepackage[noend]{algpseudocode}
\usepackage{algorithm}
\usepackage{amsmath}
\usepackage{physics}
\usepackage{multirow}
\usepackage{hyperref}
\makeatletter
\usepackage[table,xcdraw]{xcolor}
\makeatother

\title{Listen, Adapt, Better WER: Source-free Single-utterance Test-time Adaptation for Automatic Speech Recognition}
\name{Guan-Ting Lin$^1$, Shang-Wen Li$^{2*}$\thanks{$^*$Work done while working at Amazon Inc. The current affiliation is Meta AI.}, Hung-yi Lee$^1$}

\address{
  $^1$National Taiwan University, Taiwan\\
  $^2$Amazon AI, USA
\email{r10942104@ntu.edu.tw, hungyilee@ntu.edu.tw}
}
\begin{document}

\maketitle
\begin{abstract}
Although deep learning-based end-to-end Automatic Speech Recognition (ASR) has shown remarkable performance in recent years, it suffers severe performance regression on test samples drawn from different data distributions. Test-time Adaptation (TTA), previously explored in the computer vision area, aims to adapt the model trained on source domains to yield better predictions for test samples, often out-of-domain, without accessing the source data. Here, we propose the \textbf{S}ingle-\textbf{U}tterance \textbf{T}est-time \textbf{A}daptation (\textbf{SUTA\footnote{\href{https://github.com/DanielLin94144/Test-time-adaptation-ASR-SUTA}{https://github.com/DanielLin94144/Test-time-adaptation-ASR-SUTA}}}) framework for ASR, which is the first TTA study on ASR to our best knowledge. The single-utterance TTA is a more realistic setting that does not assume test data are sampled from identical distribution and does not delay on-demand inference due to pre-collection for the batch of adaptation data. SUTA consists of unsupervised objectives with an efficient adaptation strategy. Empirical results demonstrate that SUTA effectively improves the performance of the source ASR model evaluated on multiple out-of-domain target corpora and in-domain test samples.
\end{abstract}
\noindent\textbf{Index Terms}: Test-time Adaptation, Domain Shift, Speech Recognition
\section{Introduction}
Deep Learning-based Automatic Speech Recognition (ASR) models achieve impressive success, especially when samples are drawn under the independent and identical distribution (i.i.d.) assumption. However, performance degrades severely when covariate shift (i.e., distribution of test data differs from training data) happens. In real-world scenarios, such covariate-shifted test samples are ubiquitous, making ASR services unstable and unreliable. Therefore, it is critical to alleviate the adverse effect of data shifting.

Unsupervised Domain Adaptation (UDA) is a commonly-used approach that adapts the source model to the target domain without annotated data. The existing UDA approaches such as domain adversarial training \cite{sun2018domain, sun2017unsupervised}, knowledge distillation \cite{li2017large}, and self-training \cite{khurana2021unsupervised} have shown effectiveness for mitigating data shifting and improving ASR performance. 
However, they all require access to source data and sufficient target domain examples for adaptation. Such requirement imposes three main \textit{limitation} on the real-world application of UDA: (1) source data is not always available during adaptation due to the privacy/storage issues, (2) latency due to target data pre-collection and heavy computation for model adaptation, and (3) the assumption that the target domain examples come from the same distribution. 
\looseness=-1

Test-Time Adaptation (TTA) \cite{wang2020tent, liang2020we, mummadi2021test, khurana2021sita, Hu2021MixNormTA, you2021test, fleuret2021test} has recently attracted growing interest since it effectively adapts models in prediction time with little target data (a batch or even a single instance) without access to source data. 
Several studies have shown remarkable TTA effectiveness in computer vision but lack research attempts on TTA for ASR. It is worth noting that TTA in computer vision is heavily targeted on the Batch Normalization (BN) layer's adaptation by re-estimating batch statistics on target data. However, sequential models, such as ASR models, typically are not equipped with BN layers because the lengths of batched input sequences are different. Therefore, most ASR models use instance-wise Layer Normalization (LN) layer. The discrepancy in data format and model architecture also motivates us to innovate TTA tailored for ASR. 
\looseness=-1
\begin{figure}[t]
  \centering
  \includegraphics[width=\linewidth]{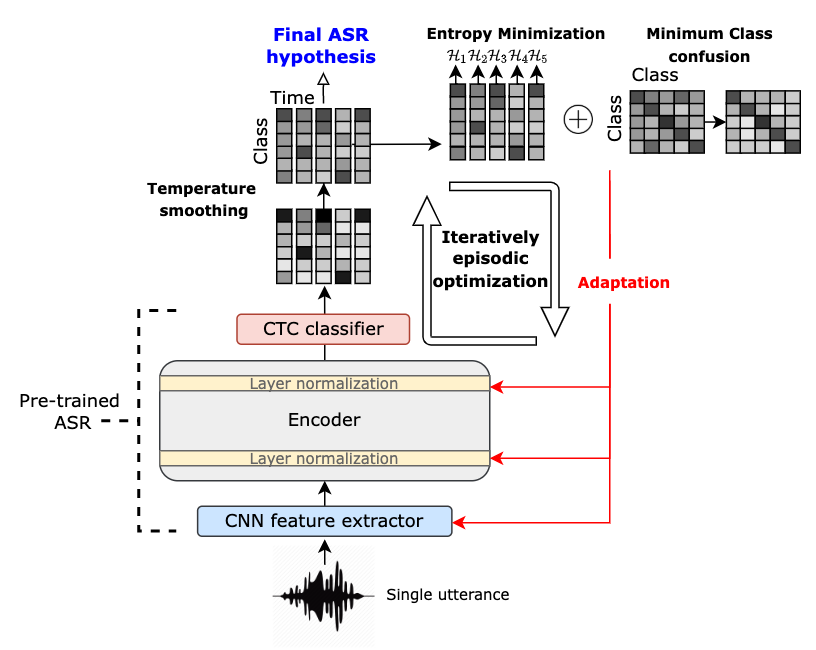}
  \vspace{-0.8cm}
  \caption{The illustration of the proposed SUTA framework. Given a single utterance, We leverage Entropy Minimization and Minimum Class Confusion as the unsupervised guidance to adapt the source ASR model in the inference time.}
  \vspace{-0.5cm}
  \label{fig:SUTA}
\end{figure}

Among different TTA variants, most methods are restricted to batch-level TTA. In other words, TTA resolves the first limitation (i.e., privacy/storage) of UDA, while the rest two (latency and target data distribution) still cause challenges.
Recently, SITA \cite{khurana2021sita} has presented a \textit{\textbf{single-instance TTA}} method, which lifts the limitation about latency and distribution since it does not require the pre-collecting batch of test samples, and it allows test samples that come from heterogeneous sources. However, SITA still focuses on computer vision problems and relies on data augmentation and statistical estimation of BN layers.

To the best of our knowledge, no existing work attempts TTA or even single-instance TTA on ASR, and there is lacking research exploring the potential of TTA in improving ASR performance. To fill the gap, in this work, we propose the \textbf{S}ingle-\textbf{U}tterance \textbf{T}est-time \textbf{A}daptation (\textbf{SUTA}) framework to improve ASR test-time performance with a single utterance. SUTA can be applied to any CTC-based end-to-end ASR model. Our method does not rely on large batch size and access to source data, which causes delayed inference and privacy issues; instead, only one testing utterance is needed for test-time adaptation in an unsupervised manner. SUTA consistently improves the source ASR model in multiple out-of-domain target corpora and in-domain test samples with little computational delay.
\section{Method}
\subsection{Problem Formulation}
We denote a trained ASR model as $g(y\mid x;\mathbf{\theta})$, parameterized by the pre-trained weight $\mathbf{\theta}$, taking utterance $x$ as input and outputs un-normalized score of vocabulary class $y$. $\mathbf{\theta}$ can be split to two parts, $\mathbf{\theta_f}$ and  $\mathbf{\theta_{a}}$. $\mathbf{\theta_f}$ is the parameters frozen at adaptation, while $\mathbf{\theta_{a}}$ is updated during the \textit{inference} time. Test corpus $D_{test}$ contains $n$ utterances $\{x_1, x_2, ..., x_n\}$.  We focus on adapting $\theta_{a}$ with a single utterance $x_i$ itself in an unsupervised manner. 
This paper uses a transformer encoder with Connectionist Temporal Classification (CTC) loss \cite{graves2006connectionist} as an ASR model, but the proposed approach is model agnostic. 
In the following, we represent the number of character classes plus one CTC blank token as $C$, and the number of time frames from the CTC classifier's output as $L$.
The model output is $\mathbf{O} \in \mathbb{R}^{L\times C}$. 

\subsection{Single-Utterance Test-time Adaptation (SUTA)}
We introduce each component of SUTA in the below sections. Figure \ref{fig:SUTA} illustrates the SUTA framework. 
\subsubsection{Entropy minimization}
Since labels are unavailable during testing, we leverage an unsupervised, entropy-based loss function for adaptation. 
Entropy Minimization (EM) intends to sharpen class distribution, which is a wide-used approach in domain adaptation and semi-supervised learning with a large amount of target data \cite{carlucci2017autodial, shu2018dirt, roy2019unsupervised, grandvalet2004semi, saito2019semi, berthelot2019mixmatch}, but not TTA. 
TENT \cite{wang2020tent} first proposes TTA by EM and adapting BN layer's parameters only with batched inputs.
Inspired by TENT, we utilize the EM objective for unsupervised test-time adaptation on model parameters with a single utterance. Since CTC blank token dominates the class distribution in $L$ frames, we exclude those frames where the CTC blank token yields the highest probability to mitigate the class-imbalanced issue. $\mathcal{L}_{em}$ can be calculated as
\begin{equation}
    \mathcal{L}_{em} = \frac{1}{L} \sum_{i=1}^L \mathcal{H}_i = - \frac{1}{L} \sum_{i=1}^L \sum_{j=1}^C \mathbf{P_{ij}} \log \mathbf{P_{ij}} , 
\end{equation}
where $\mathcal{H}_i$ is the entropy at $i$-th frame, and $\mathbf{P_{ij}}$ is the probability of the $j$-th class in the $i$-th frame. 
\subsubsection{Minimum class confusion}
In addition to the EM objective, Minimum Class Confusion (MCC) objective is an alternative adjusting model parameters by reducing the correlation between different classes. This objective is adopted by \cite{jin2020minimum, chen2019domain, vu2019advent} for domain adaptation, and \cite{you2021test} for batched test-time adaptation in computer vision. The equation of MCC loss is
\begin{algorithm}[]
\caption{SUTA Algorithm}\label{alg:main}
\begin{algorithmic}[1]
\State $C:$ number of classes, $L:$ number of frames, $T:$ temperature, $\mathbf{\theta_{f}}:$ frozen weight, $\mathbf{\theta_{a}}:$ adaptation weight, $x:$ an utterance in test set $D_{test}$
\For{$t=1$ to $N$}
    
    \State $\mathbf{O} = g(y\mid x; \mathbf{\theta_{f}}, \mathbf{\theta_{a}})$
    \State $\mathbf{P_{\cdot j}} = \frac{\exp(\mathbf{O_{\cdot j}}/T)}{\sum_{j'= 1}^C \exp(\mathbf{O_{\cdot j'}}/T)} , \forall j \in C$
    
    \State $\mathcal{L}_{em} = - \frac{1}{L} \sum_{i=1}^L \sum_{j=1}^C \mathbf{P_{ij}} \log \mathbf{P_{ij}}$
    \State $\mathcal{L}_{mcc} = \sum_{j=1}^C \sum_{j'\neq j}^C \mathbf{P_{\cdot j}^\top P_{\cdot j'}}$
    \State $\mathcal{L} =  \alpha \mathcal{L}_{em} + (1-\alpha) \mathcal{L}_{mcc}$ 
    \State $\mathbf{\theta^{t+1}_{a}} = $ \textrm{Optimizer}$(\mathbf{\theta^t_{a}}, \mathcal{L})$
\EndFor
\State Output: Decode $g(y\mid x; \mathbf{\theta_{f}}, \mathbf{\theta^{N}_{a}})$.
\end{algorithmic}
\end{algorithm}
\vspace{-0.3cm}
\begin{equation}
    \mathcal{L}_{mcc} = \sum_{j=1}^C \sum_{j'\neq j}^C \mathbf{P_{\cdot j}^\top P_{\cdot j'}} , 
\end{equation}
where $\mathbf{P_{\cdot j}} \in \mathbb{R}^{L}$ denotes the probabilities of the $j$-th class of the $L$ frames.  
We minimize the class correlation between pairs of class $j$ and $j'$ ($j \neq j'$) by only penalizing non-diagonal values on the class confusion matrix. Actually, \cite{chen2019domain} showed that the MCC objective is similar to EM, but the gradient is different. 
\looseness=-1
\subsubsection{Temperature smoothing}
However, we discovered that naively TTA with $\mathcal{L}_{em}$ and  $\mathcal{L}_{mcc}$ yield minor performance improvement empirically (Table \ref{tab:ablation} $T=1$ row). These results can be explained since entropy loss is small for the confident predictions, so the EM objective cannot gain guidance from those confident frames. Instead, the $\mathcal{L}_{em}$ mainly attributes to the uncertain frames, which may provide unreliable update direction. Very recently, \cite{mummadi2021test} also unrevealed gradient vanishing problem for high confident predictions. To cope with this issue, we use the temperature scaling method to smooth probability distribution, keeping the influence of high-confident frames. The temperature smoothing can also alleviate the negative effects of over-confident prediction for MCC objectives. We smooth the output distribution as
\begin{equation}
   \mathbf{P_{\cdot j}} = \frac{\exp(\mathbf{O_{\cdot j}}/T)}{\sum_{j'= 1}^C \exp(\mathbf{O_{\cdot j'}}/T)} ,
\end{equation}
where $\mathbf{O_{\cdot j}}$ is the output logits of $j$-th class for all a time frame. $T$ is larger than 1 for flattening the probability distribution. At every iteration, the smoothed distribution is used for loss calculation in equation (1) and (2).
\subsubsection{Training objective}
To prevent models from overfitting on one unsupervised objective, both $\mathcal{L}_{em}$ and $\mathcal{L}_{mcc}$ losses are optimized for adaptation. $\alpha$ is the hyper-parameter for weighted-sum two loss. Overall loss function can be written as below:
\begin{equation}
    \mathcal{L} =  \alpha \mathcal{L}_{em} + (1-\alpha) \mathcal{L}_{mcc}
\end{equation}
The overview of SUTA is in Algorithm \ref{alg:main}. For each utterance $x$, starting from the original weights of source ASR model, we forward $x$ to obtain model output $\mathbf{O}$, temperature smoothing the distribution, and train the adaptation parameter $\mathbf{\theta_{a}}$ by minimizing the loss function $\mathcal{L}$. After \textit{N} iterations, the adapted ASR model $g(y\mid x; \mathbf{\theta_{f}}, \mathbf{\theta^{N}_{a}})$ is used for final inference.
\begin{table*}[t]
\centering
\caption{Word Error Rate (\%) for different corpora and methods. All WERs are measured without decoding with Language models. State-of-the-art (SOTA) performances are from \cite{chung2021w2v} for LS, \cite{nttchime3} for CH, \cite{zhou2020rwth} for TD, and \cite{speechbrain} for CV. The dashed line ``-" means there is no performance reported by prior works.}
\vspace{-0.3cm}
\begin{tabular}{cccccccc}
\toprule
\multirow{2}{*}{\begin{tabular}[c]{@{}c@{}}\textbf{Performance reference for source ASR model}\\ \textit{wo/ adaptation}\end{tabular}}  & \multicolumn{3}{c}{LS test-o + $\delta$}                                              & \multirow{2}{*}{CH} & \multirow{2}{*}{CV} & \multirow{2}{*}{TD} \\
                        &                            \multicolumn{1}{c|} 0 & \multicolumn{1}{c|}{0.005} & 0.01 &                        &                       &                      \\ \midrule 
SOTA (trained on target dataset)                              & 2.5                           & -                     & -         & 5.8             & 15.4               & 5.6            \\ 
RASR \cite{likhomanenko2020rethinking} (trained on LS)                               & 6.8                       & -                    & -       & -                 & 29.9                & 13.0              \\ 

\midrule
\midrule
\textbf{TTA method} &                         &                     &         &             &            &               \\ \midrule

(1) Our source ASR model \cite{baevski2020wav2vec} (trained on LS \textit{wo/ adaptation})                             & 8.6                            & 13.9                     & 24.4        & 31.2                & 36.8              & 13.2               \\ 
(1) + SDPL               &             8.3                    &           13.1                 & 23.1         &        30.4                &        36.3               &      12.8                \\
(1) + SUTA                  & \textbf{7.3}                            & \textbf{10.9}                      & \textbf{16.7}         & \textbf{25.0}                 & \textbf{31.2}                 & \textbf{11.9}                \\ \bottomrule

\label{tab:main}
\end{tabular}
\vspace{-0.8cm}
\end{table*}
\section{Experiments}
\subsection{Source ASR model}
We use open-sourced CTC-based ASR model, Wav2vec 2.0-base CTC model \cite{baevski2020wav2vec}, as our source ASR model. This model is fine-tuned on Librispeech 960-hour\footnote{Checkpoint of the trained ASR model is open-sourced in https://huggingface.co/facebook/wav2vec2-base-960h} and obtains 3.4 and 8.6 Word Error Rate (WER) on Librispeech test-clean and test-other set. Since this model is trained on Librispeech, we regard Librispeech as the source domain. The architecture of the Wav2vec 2.0-base CTC ASR model is composed of three parts: a convolutional neural network-based feature extractor, a 12-layer transformer encoder, and a linear CTC classifier. 
\subsection{Datasets}
To evaluate the adaptation capabilities of the proposed SUTA approach, we examine the test-time performance in the following target domains: \textbf{(1) Librispeech (LS)} \cite{panayotov2015librispeech} contains read speech from audiobook in 16kHz. Although Librispeech test data is in-domain for our ASR source model, we want to verify whether SUTA retains performance for in-domain test samples. Furthermore, we inject additive Gaussian noises for different amplitudes ($\delta=\{0.005, 0.01\}$) on the test-other set to create covariate shift, testing SUTA's capability to mitigate Gaussian noises. \textbf{(2) CHiME-3 (CH)} \cite{barker2017third} is a noisy version of WSJ corpus with artificial and real-world environmental noises at 16kHz. We utilize the official enhanced evaluation set \texttt{et05} to examine SUTA's robustness in noisy acoustic conditions. \textbf{(3) Common voice (CV)} \cite{ardila-etal-2020-common} is a crowdsourcing project that is supported by volunteers to read Wikipedia sentences and record samples at 48kHz. We re-sample the sampling rate to 16kHz to match the training condition of the source ASR model. Test set from the \texttt{En-June-22nd-2020} release version is utilized. \textbf{(4) TEDLIUM-v3 (TD)} \cite{hernandez2018ted} consists of oratory speech based on TED conference videos. The audio quality is clean and stored at 16kHz. We use the official test set for experiments. For all these datasets, transcripts are pre-processed by upper-casing letters and removing punctuation except for apostrophes following \cite{likhomanenko2020rethinking}.
\subsection{Baseline TTA methods}
Since there is no existing single-utterance TTA study on ASR, here we propose a naive pseudo labeling approach named Single-utterance dynamic pseudo labeling (\textbf{SDPL}) as the baseline method, which uses the source ASR model to predict the pseudo label of the utterance and adapt the model by minimizing CTC loss. The pseudo label is generated by greedy decoding on CTC output and refined at each iteration dynamically. SDPL does not consider the uncertainty of the distribution due to the nature of greedy decoding. 
\looseness=-1
\subsection{Implementation details}
The hyperparameters are chosen from LS test-o with noises ($\epsilon=0.01$), and we use the best-found setup for all other target corpora. We explore different settings of trainable adaptation parameters $\mathbf{\theta_{a}}$ in SUTA. For simplicity, we denote the parameters of layer norm transformation as \texttt{LN}, feature extractor as \texttt{Feat}, and entire ASR model as \texttt{All}. The optimizer is AdamW, and the learning rate is searched from $10^{-4}$ to $10^{-6}$. The best-found learning rate is different for each setup of trainable parameters. Specifically, the best learning rate is $2e^{-4}$, $2e^{-5}$, and $1e^{-6}$ for \texttt{LN}, \texttt{LN+Feat}, and \texttt{All}. It is intuitive as the more parameters are relaxed for adaptation, the lower the learning rate is. We noted that SDPL could only improve WER when adapting with \texttt{LN}; more parameters cause drastic WER deterioration (probably due to errors from greedy pseudo label generation). Thus, we only report results of LN for SDPL in Section \ref{sec:4}.
\looseness=-1

For default setup in all SUTA experiments, $\alpha$ is $0.3$, the number of adaptation iterations \textit{iter} is 10, and $\mathbf{\theta_{a}}$ is \texttt{LN+Feat}. Experiments are run on Nvidia Geforce RTX 3090 GPU. Adaptation speed is about 0.115 second for a 1 second utterance for 10-steps adaptation. 

\subsection{Results}
\label{sec:4}
Table \ref{tab:main} summarizes our experiment results. We list the state-of-the-art performances of models trained and evaluated on each dataset as the topline WER in the first row. We then show the RASR's \cite{likhomanenko2020rethinking} results as a baseline where source (training) and target (evaluation) domains mismatch. RASR is a strong baseline, which leverages a much bigger network (36 transformer layers) and is trained with data augmentation (SpecAugment \cite{park2019specaugment}). We also present results for our source ASR model, Wav2vec 2.0-based CTC model trained with source domain data (LS). Lastly, we show the performance of two single-utterance adaptation techniques, SDPL and our proposed SUTA, on top of our source model. SOTA results are unsurprisingly better than others since the model is trained and evaluated without mismatch. SUTA outperforms our source model and the SDPL baseline consistently. Moreover, our proposed approach yields comparable results to RASR with smaller networks and a training/adaption process, suggesting SUTA's efficacy. Noting that we can also adopt SUTA on RASR's models to improve test-time performance, here the results of RASR are just for showing a strong baseline without adaptation. 

We dive deep into the results. For the in-domain evaluation (i.e., LS without additive noise), we observe that the pseudo labeling baseline (SDPL) slightly improves WER on top of the source ASR model (c.f., 8.3 vs. 8.6). SUTA further improves the WER to 7.3, which is surprising that TTA can even enhance the test-time performance for in-domain samples. With the target domain drifting from source by adding Gaussian noises, WER degrades from 8.6 to 13.9 ($\delta=0.005$) and 24.4 ($\delta=0.01$) for our source model. SDPL slightly improves WER, while SUTA drastically reduces WER for 3.9 ($\delta=0.005$) and 7.7 ($\delta=0.01$), suggesting that SUTA can improve the robustness of ASR when Gaussian noises are present.

Besides synthesizing the drifted target domain with LS, SUTA mitigates mismatch from multiple real-world audio corpora. For noisy speech in CH, SUTA successfully boosts WER from 31.2 to 25.0, whereas SDPL can only reduce WER to 30.4. Similar results can be found with the crowdsourcing recording in CV and the clean oratory speech in TD. SUTA improves WER from 36.8 to 31.2 for CV and from 13.2 to 11.9 for TD, outperforming SDPL significantly. 
\vspace{-0.2cm}
\begin{table}[t]
\centering
\caption{Ablation study on CH evaluation set. Besides WER, we report the relative WER reductions (WERR) from Baseline to different ablation settings.}
\vspace{-0.3cm}
\label{tab:ablation}
\begin{tabular}{lll}
\toprule
S\textbf{pecification}  & \textbf{WER (\%)} & \textbf{WERR (\%)} \\ \midrule

Baseline (unadapted)              &  31.2   &   0.0   \\ \midrule
\rowcolor[HTML]{EFEFEF} 
\multicolumn{3}{l}{\textit{SUTA best config.}}  \\
$\alpha$=0.3, LN+Feat, $T$=2.5   &  25.0   &  19.9     \\
\midrule
\rowcolor[HTML]{EFEFEF} 
\multicolumn{3}{l}{\textit{Weighted loss coefficient ($\alpha$})}                                                                       \\
$\alpha$=1.0                                                                           &  25.5   &   18.3   \\
$\alpha$=0.7                                                                           &  25.3   &   18.9   \\
$\alpha$=0.5                                                                           &  25.1   &   19.6   \\
$\alpha$=0.0                                                                           &  25.4   &   18.6   \\
\midrule
\rowcolor[HTML]{EFEFEF} 
\multicolumn{3}{l}{\textit{Temperature smoothing (T)}}                                                             \\
$T$=1.0                                                                                  &  29.9   &  4.2    \\
$T$=1.5                                                                                  & 26.7   &  14.2    \\
$T$=2.0                                                                                  &  25.6   &  17.9    \\
$T$=3.0                                                                                  &  25.5   &  18.3   \\
\midrule
\rowcolor[HTML]{EFEFEF} 
\textit{Trainable parameters ($\mathbf{\theta_{a}}$)}                                                               &     &      \\
LN                                                                                     &  28.8   &   7.7   \\
Feat                                                                                     &  25.1   &  19.6    \\
All                                                                                    &  26.7   &   14.4   \\ 
\rowcolor[HTML]{EFEFEF} 
\bottomrule
\end{tabular}
\vspace{-0.5cm}
\end{table}
\subsection{Discussion and analysis}
We investigate the importance of different components in SUTA, including the weighted sum coefficient $\alpha$ of the loss function, temperature values, and trainable parameters. Due to space limitations, we only present an analysis on CH.\\
\textbf{Ablation study}: Ablation studies are summarized in Table \ref{tab:ablation}. Both EM and MCC objectives improve the performance individually, according to row $\alpha=1.0$ and $\alpha=0.0$, and weighted-sum combination further enhances relative WER reduction (WERR) from 18.3 / 18.6 to 19.9 when $\alpha=0.3$. The results indicate that using multiple objectives can improve the TTA, preventing overfitting on a single loss. Besides, the contribution of temperature smoothing is significant, reducing from 29.9 (T = 1.0) to 25.0 (T= 2.5) WER. 

\begin{figure}[t]
  \centering
  \vspace{-0.3cm}
  \includegraphics[width=0.7\linewidth]{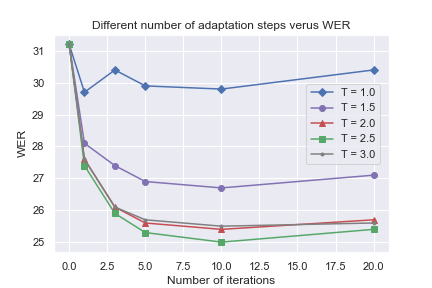}
  \vspace{-0.4cm}
  \caption{Different temperature values (T) for temperature smoothing at different number of iterations on CHiME-3 evaluation set.}
  \vspace{-0.55cm}
  \label{fig:step}
\end{figure}

\begin{figure}[t]
  \centering
  \includegraphics[width=0.7\linewidth]{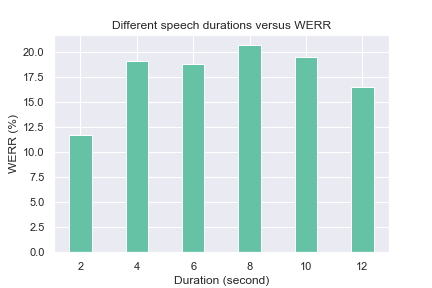}
  \vspace{-0.35cm}
  \caption{The analysis of how utterance duration influence the SUTA effectiveness on CHiME-3 evaluation set.}
  \vspace{-0.7cm}
  \label{fig:duratiom}
\end{figure}

Choosing appropriate trainable parameters is also crucial to SUTA. There is 7.7 WERR by solely adapting the layer normalization (\texttt{LN}). If we fine-tune the normalization layer and the feature extractor (\texttt{LN+Feat}), we can further increase performance gain to $19.9$ WERR. Nevertheless, relaxing too many parameters for single-utterance fine-tuning causes slight performance degradation, as indicated by the result for \texttt{All}, which only yields 14.4 WERR.\\
\textbf{Iteration step}: We analyze the different number of model adaptation steps and their impact on performance. We explore  1, 3, 5, 10, and 20 for the number of iterations. Figure \ref{fig:step} demonstrates that the WER is consistently reduced before 10 steps; however, after 10 steps, the improvement saturates, particularly when not adopting temperature smoothing ($T=1$). On the other hand, utilizing temperature smoothing ($T>1$) stabilizes adaptation, mitigates the performance degradation in the later iterations, and enhances overall effectiveness.\\
\textbf{Utterance length}: When the length of input audio is short, it only contains one or two words. In this case, the output prediction merely includes a small number of classes and may be prone to class collapse when entropy is minimized. We investigate the relationship between utterance length and WERR in Figure \ref{fig:duratiom}. The result indicates that the WERR of short utterances (less than 2 seconds) is about 12 \% WERR, which is much lower than the around 17.5 \% WERR observed in typical utterances (those longer than 2 seconds). Although the performance gain of short audio is smaller than the normal one, we do not observe any adverse effect, and there is still WER improvement. 
\section{Conclusion}
In this work, we propose the first source-free \textbf{S}ingle-\textbf{U}tterance \textbf{T}est-time \textbf{A}daptation (SUTA) framework on ASR, which can efficiently adapt CTC-based ASR models given one target utterance in the inference time. Specifically, we use entropy minimization and minimum class confusion objectives for TTA and further improve TTA by temperature smoothing. Experimental results demonstrate that SUTA effectively reduces the source ASR model's WER on several out-of-domain corpora, even enhancing the performance of in-domain test samples. We plan to design TTA on sequence-to-sequence ASR models in the future, improving performance on short utterances. 
\bibliographystyle{IEEEtran}

\bibliography{mybib}

\end{document}